\documentclass{iauJDSS}
\usepackage{graphicx}

\title[Are jets rotating?] 
{Are jets rotating at the launching?}

\author[N. Soker]   
{Noam Soker$^1$}

\affiliation{$^1$ Department of Physics, Technion$-$Israel
Institute of Technology, Haifa 32000 Israel; soker@physics.technion.ac.il.}

\pubyear{2010}
\jname{Astrophysical Outflows and Associated Accretion Phenomena}
\editors{I. F Corbett, E. de Gouveia Dal Pino, A. Raga, eds.}
\begin{document}

\maketitle

\begin{abstract}
I argue that the Doppler shift asymmetries observed in some young stellar object (YSO) jets
result from the interaction of the jets with the circumstellar gas, rather than from
jets' rotation. The jets do rotate, but at a velocity much below claimed values.
During the meeting I carefully examined new claims, and found problems with the
claimed jets' rotation.
I will challenge any future observation that will claim to detect jet rotation in YSOs
that requires the jets (and not a wind) to be launched from radii much larger than the
accreting stellar radius.
I conclude that the most likely jets' launching mechanism involves
a very efficient dynamo in the inner part of the accretion disk, with a jets' launching mechanism
that is similar to solar flares (coronal mass ejection).
\end{abstract}

\section{Introduction}

In previous papers (Soker 2005, 2007a)  I proposed that the interaction of the jets with
a twisted-tilted (wrapped) accretion disk can form the asymmetry in the jets'
  line of sight velocity profiles as observed in some YSOs (e.g. Bacciotti et al. 2002).
The claim that the observations of asymmetric Doppler shifts do not support
jet rotation in YSOs was strengthened by the numerical simulations of Cerqueira et al. (2006).
They assumed a precessing jet whose ejection velocity changes periodically with a period
equals to the precession period.
Practically, the dependance of the jet's expansion velocity on direction around the
symmetry axis leads to the same effect as the model of Soker (2005).
Whereas in Soker (2005) the physical process behind this velocity profile is an
interaction with the material in the jet's surroundings, Cerqueira et al. (2006) give no
justification for the periodic variation of the jet's ejection speed.
As far as comparison with observation is considered, it is hard to
distinguish between the model of jet interaction with its surrounding (Soker 2005),
and the periodic jet's speed of Cerqueira et al. (2006).

\section{Problems with claimed jets' rotation}

To demonstrate the problems with the argued jet rotation, I will examine two new claims.

 After the publication of my earlier papers Zapata et al. (2009) argued for
a rotating molecular jet in Ori-S6.
I find four problems with this case. More detail can be found in my presentation
at the meeting:
\newline
http://iaujd-outflows.blogspot.com/2008/10/scientific-program.html
\newline
(1) In some regions the red and blue shifted components overlap. This is against expectation
if the red-blue shifted components are due to jets' rotation.
(2) In some regions the blue and red shifted components are disconnected.
As each jet is one entity, this is against expectation if the Doppler shifts are due to jets' rotation.
(3) Using the rotation interpretation at the edge of $ 30^{\prime \prime}$
across the disk, gives a jet's foot-point of $300~$AU.
This is larger than the size of the accretion disk given by the same authors for this object.
(4) The ring that supposedly feeds the accretion disk and the jets, rotates
in opposite sense to that of the claimed jets' rotation.  As Zapata et al. write:
``The sense of rotation of the circumbinary ring is nearly opposite to that of jet and
       outflow, and the jet leaves the system under an angle of $45^{\circ}$ with the ring plane.''
I note that a tilted jet can lead to the asymmetric red-blue shift, as in the model
I proposed in 2005.

During the meeting, I was challenged to account for a very recent claim
of a possible jets' rotation in HH~211 (Lee et al. 2009, 2007).
I find two problems with the tentative claimed jets' rotation
(I elaborate on these points in the appendix in the astro-ph version of this paper).
(1) The blue and red components exchange sides.
Namely, the velocity plots do not give a clear sense of asymmetry, and hence no unique sense
of rotation. The same effect is seen in the velocity maps of HH~212 (Lee et al. 2008).
(2) The accretion disk cannot supply the required anergy and angular momentum if
the rotation is real.

My conclusion is that these types of observations give peaks in emission that
show different Doppler shifts.
By pure fluctuations, these might mimic rotation in some places.
In same cases the sense of the fluctuations will give rotation in the same
sense as that of the accretion disk. In other cases the sense will be in an
opposite direction to that of the disk, and in some cases just zero rotation will be deduced.
The inferred rotation is due to fluctuations that by chance can mimic rotation.

\section{The launching mechanism}

The talks and discussions during the meeting strengthened my view that the
launching mechanism involves reconnection of magnetic field lines.
Reconnection can occur between the stellar and the disk magnetic fields
(e.g., de Gouveia dal Pino \& Lazarian 2005; de Gouveia dal Pino et al. 2009),
or reconnection of the disk magnetic field (Soker 2007b).
Laor \& Behar (2008) show that the ratio of radio luminosity  to X-ray luminosity
has similar values in magnetically active stars and in many accreting objects, up
to radio quiet quasars. Based on this correlation I prefer the following conclusion
(Soker 2007b; Soker \& Vrtilek 2009):
There is a very efficient dynamo in the inner part of
the accretion disk, with a jets' launching mechanism
that is similar to solar flares (coronal mass ejection).


\newpage
\section{Appendix (astro-ph version only): Do the jets in HH211 rotate? }

I took the challenge raised during my talk, and looked at the papers by Lee et al. on HH211
(Lee et al. 2007, 2009). There are two problematic points.

\subsection {Velocity Maps}

In figure 11 of Lee et al. (2009) they show a 1 km/sec velocity map.
They claim the red component peaks on one side, while the blue on the other.
However, it seems that the blue and red components exchange sides, as marked on
the figure below.
From what I can tell from other figures in their second paper,
the velocity plots do not give a clear sense of asymmetry.
The same feature appears in the velocity maps of HH~212 presented by
Lee et al. (2008). In their figs. 4 and 6 the sign of the velocity gradient changes
between different locations. In their fig. 1b the red and blue components exchange sides.
Here as well, fluctuations seems to cause these variations.
\begin{figure}
\hskip -3.8 cm
\includegraphics[scale=0.91]{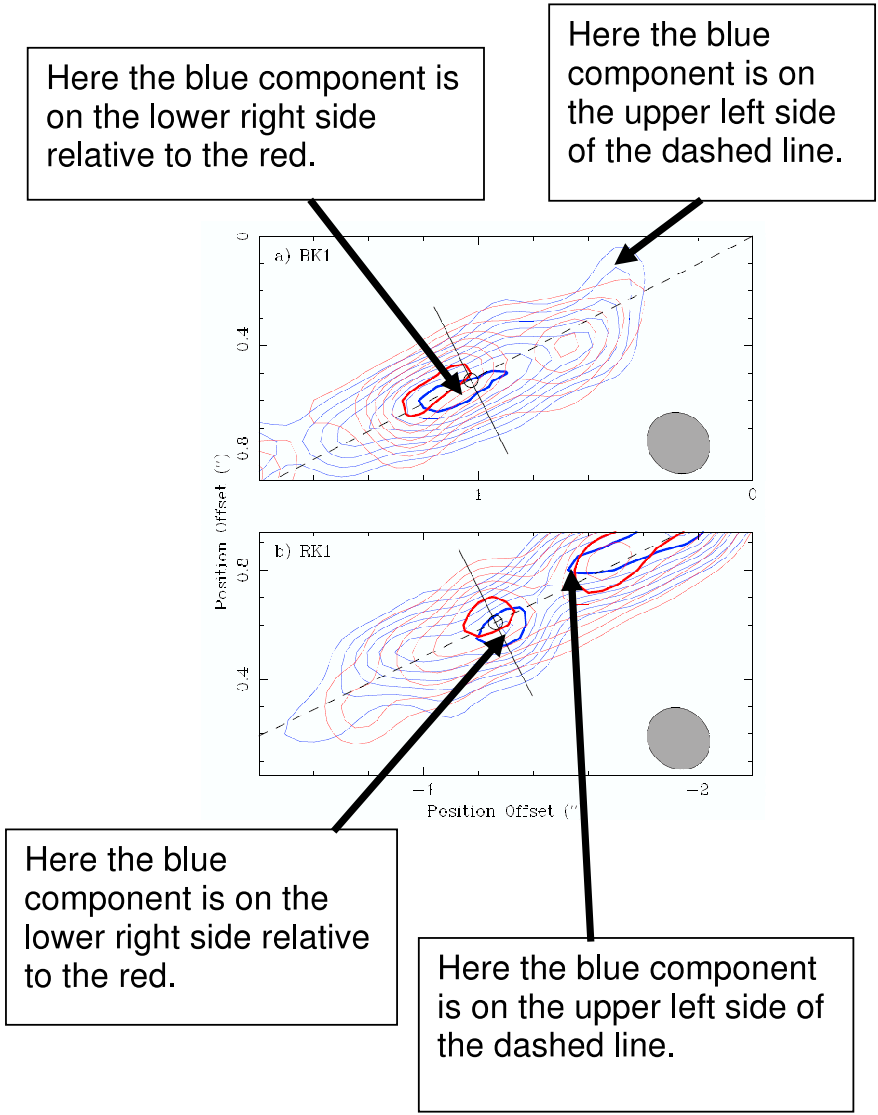}
\vskip -2.5 cm
\caption{The velocity contour map from figure 11 of Lee et al. (2009).
Marked are the changes in the sense of rotation. }
\label{fig1}
\end{figure}

\subsection {The foot-point}

In the first paper they derive the two-sided mass-loss rate of the jets
\begin{equation}
\dot M_{\rm 2j}=(0.7-2.8) \times 10^{-6} M_\odot ~{\rm yr}^{-1}.
\label{eq1}
\end{equation}
For the accretion rate they give the following values for the first and second
paper, respectively
\begin{equation}
\dot M_{\rm acc-f}=8 \times 10^{-6} M_\odot ~{\rm yr}^{-1}; \qquad
\dot M_{\rm acc-s}=(2.5-5) \times 10^{-6} M_\odot ~{\rm yr}^{-1}.
\label{eq23}
\end{equation}
Taken all these values, the ejection in the jets to accretion ratio is
\begin{equation}
\eta_{\rm f} \equiv \frac{\dot M_{\rm acc-f}}{\dot M_{\rm 2j}} \simeq 0.09-0.36;
\qquad
\eta_{\rm s} \equiv \frac{\dot M_{\rm acc-s}}{\dot M_{\rm 2j}} \simeq 0.2-0.6,
\label{eq4}
\end{equation}
by the first and second paper, respectively.

A jet velocity of  $v_j=170~ {\rm km ~ s}^{-1}$  from their star of mass $M=0.05 M_\odot$,
and the above ratio $\eta$, require the foot point to be at radius $r_0$
determined from energy conservation
\begin{equation}
\frac{1}{2} \frac{GM}{r_0}\dot M_{\rm acc}
= \frac{1}{2} \dot M_{\rm 2j} v_j^2.
\label{eq6}
\end{equation}
This gives
\begin{equation}
r_0= \frac{GM}{v_j^2} \eta^{-1} = 3.3
\left( \frac{\eta}{0.1} \right)^{-1} R_\odot.
\label{eq7}
\end{equation}
The specific angular momentum that the accreted mass can supply to the jet is
\begin{equation}
j_m = \frac {\sqrt{GMr_0}}{\eta} =
8.3 \left( \frac{\eta}{0.1} \right)^{-1}
\left( \frac{r_0}{3.3~{\rm AU}} \right)^{-1}
=8.3 \left( \frac{\eta}{0.1} \right)^{-2} {\rm km ~s}^{-1}~{\rm AU}.
\label{eq8}
\end{equation}

The claimed observed value is $j_m =5 ~{\rm km ~s}^{-1}~{\rm AU}$.
Namely, the values of $\eta$ is constraint to be $\eta < 0.15$.
It cannot be as large as 0.3 as it is according to Lee et al. (2009).
As I do not expect that the accreted mass will transfer angular momentum and
energy at 100\% efficiency to the ejected gas
(namely, its rotation speed will not drop to zero),
the constraint on $\eta$ is stronger even, $\eta < 0.1$.


\begin{thebibliography}{}


\bibitem[Bacciotti (2002)]{Bacciotti}
Bacciotti, F., Ray, T. P., Mundt, R., Eisl\"offel, J., \& Solf, Jo. \textit{ApJ}, 576, 222 (B2002)

\bibitem[Cerqueira et al. (2006)]{Cerqueira06}
Cerqueira, A. H., Velazquez, P. F., Raga, A. C., Vasconcelos, M. J., \& de Colle, F. 2006,
         \textit{A\&A}, 448, 231

\bibitem[de Gouveia dal Pino \& Lazarian (2005)]{dal Pino05}
de Gouveia dal Pino, E. M., \& Lazarian, A. 2005, \textit{A\&A}, 441, 845

\bibitem[de Gouveia Dal Pino (2009)]{DalPino09}
de Gouveia Dal Pino, E. M., Piovezan, P., Kadowaki, L., Kowal, G., \& Lazarian, A.,
  this proceedings  (IUA JD 7 at the XXVIIth IAU General Assembly)

\bibitem[Laor \& Behar (2008)]{Laor08}
  Laor, A., \& Behar, E. 2008, \textit{MNRAS}, 390, 847


\bibitem[Lee et al. (2007)]{Lee07}
    Lee, C.-F., et al.   2007, \textit{ApJ}, 670, 1188

\bibitem[Lee et al. (2008)]{Lee08}
    Lee, C.-F., et al.   2008, \textit{ApJ}, 685, 1026

\bibitem[Lee et al. (2009)]{Lee09}
    Lee, C.-F., et al.   2009. \textit{ApJ}, 699, 1584

\bibitem[Soker (2005)]{soker2005}
       {Soker, N.} 2005, \textit{A\&A}, 435, 125

\bibitem[Soker (2007a)]{soker2007a} {Soker, N.} 2007a, astro-ph/0703474

\bibitem[Soker (2007b)]{soker2007b}
 Soker, N. 2007b, IAUS, 243, 195 (arXiv:0706.4241).

\bibitem[Soker \& Vrtilek (2009)]{sokerv09}
 Soker, N., \& Vrtilek, S. D.  2009, arXiv:0904.0681

\bibitem[Zapata et al. (2009)]{Zapata}
 Zapata, L. A., Schmid-Burgk, J., Muders, D., Schilke, P., Menten, K., \& Guesten, R.
    2009, \textit{A\&A} in press


\end{thebibliography}
\end{document}